\documentclass[superscriptaddress,showpacs,a4paper,twocolumn]{revtex4}
\usepackage{graphicx}
\usepackage[latin1]{inputenc}

\newcommand{\EVRY}{Universit\'e d'Evry-Val d'Essonne, Boulevard François Mitterrand, 91025 Evry Cedex, France}
\newcommand{\LKB}{Laboratoire Kastler Brossel, UPMC-Paris 6, ENS, CNRS ; Case 74, 4 place Jussieu, 75005 Paris, France}

\begin{document}
\title{Why three-body physics does not solve the proton radius puzzle}

\author{Jean-Philippe Karr}
\email{karr@spectro.jussieu.fr}
\affiliation{\LKB}
\affiliation{\EVRY}

\author{Laurent Hilico}
\affiliation{\LKB}
\affiliation{\EVRY}

\date{\today}
\begin{abstract}
The possible involvement of weakly bound three-body systems in the muonic hydrogen spectroscopy experiment~\cite{pohl2010}, which could resolve the current discrepancy between determinations of the proton radius, is investigated. Using variational calculations with complex coordinate rotation, it is shown that the $p\mu e$ ion, which was recently proposed as a possible candidate~\cite{jentschura2011a}, has no resonant states in the energy region of interest. QED level shifts are included phenomenologically by including a Yukawa potential in the three-body Coulomb Hamiltonian before diagonalization. It is also shown that the $pp\mu$ molecular ion cannot play any role in the observed line.
\end{abstract}
\maketitle

\textbf{Introduction.} The recent measurement of the Lamb shift in muonic hydrogen~\cite{pohl2010}, which resulted in a new determination of the proton rms charge radius, has given rise to an abundant literature attempting to resolve the 4-percent discrepancy with its current CODATA value~\cite{codata}, mainly obtained from electron-proton (e-p) scattering and hydrogen (H) spectroscopy. The discrepancy with e-p scattering measurements~\cite{bernauer2010} may be less severe than the initially announced 5 standard deviations due to a possible underestimation of error bars in e-p data analysis~\cite{belushkin2007}. However, no agreement with H spectroscopic data~\cite{schwob1999} could be obtained unless some unknown systematic shift was discovered in {\em all} these experiments. An attempt to solve the discrepancy by re-evaluating the third Zemach moment of the proton~\cite{derujula2010} was shown to go against e-p scattering data~\cite{cloet2011}. Reconsideration of the assumptions on form factors used to extract the proton radius only led to a slight increase of the uncertainty~\cite{carroll2011}. Much effort has been focused on checking and improving the Lamb shift calculations used to extract the proton radius (see e.g.~\cite{jentschura2011a,borie2012} and references therein, as well as~\cite{yerokhin2011}). In particular, the polarizability of the proton is a subject of hot debates~\cite{carlson2011a}, but no consensus has been reached yet regarding claims that this contribution may be able to solve the proton size puzzle. The muonic hydrogen result also triggered a search for new physics, which is however highly constrained by many low-energy data~\cite{jaeckel2010}.

So far, no possible experimental artifacts have been proposed, with the notable exception of ~\cite{jentschura2011a}. In that reference, Jentschura suggested that $\mu p$ atoms formed in the 2S state might bind an electron present in the gas target to form a $(p \mu e)^-$ ion. Since the muon orbit is close to the proton, the proton charge is shielded from the outer electron, but attractive dipole interactions with the $\mu p$(2S) core could result in a quasibound (or resonant) state slightly below the 2S threshold. Jentschura gave an order-of-magnitude estimate of the associated frequency shift, which was indeed compatible with the observed discrepancy $\Delta E_{exp} - \Delta E_{th} = +0.31$~meV between experiment and theory. Among the many proposed solutions to the proton radius puzzle, this appears to be one of the last unexplored paths, and definitely deserves further study. The purpose of Section~\ref{sec-pmue} is to give a conclusive -eventually negative- answer on the viability of the $p \mu e$ hypothesis.

Other candidates to form such quasibound states with a $\mu p$ atom can also be considered. While the possibility of forming $(p \mu \mu)^-$ ions can be safely ignored -if only because of the low number of muons in the experiment chamber, it is known that $(pp\mu)^+$ molecules can be formed in collisions of excited $\mu p$(2S) atoms with hydrogen molecules (see e.g.~\cite{wallenius2001} and references therein). In this process, the favored molecular states lie $\sim 10-20$~eV below the $n=2$ dissociation threshold, and their Coulomb lifetime is much too short to account for the observed transitions~\cite{kilic2004,lindroth2003}. However, it is worth reconsidering the possible role of more weakly bound states which can have long lifetimes. In Section~\ref{sec-ppmu}, we discuss the relevant portion of the spectrum of $pp\mu$ and show that none of the existing states has the required characteristics to match the experimental data.

\section{The $p\mu e$ hypothesis} \label{sec-pmue}

\subsection{Coulomb Schr\"odinger Hamiltonian}

A first insight into the physics of the $p\mu e$ system is provided by the adiabatic approximation: the ''slow'' outer electron experiences the attractive dipole potential created by the $p\mu (n=2)$ core. Neglecting (in a first step) the 2S-2P Lamb shift, the $n=2$ degenerate level has a nonzero static dipole $A = 3 m_e/m_{r}$ in atomic units, where $m_r$ is the $p\mu$ reduced mass. The long-range behavior of the potential curve is then $-A/R^2$, and it has been known for a long time~\cite{morse} that in this kind of potential, a particle of mass $m$ has an infinite series of bound states with exponentially decreasing binding energies, provided $A>A_c = m_e/8m$. This translates into the condition $m > m_r/24 \simeq 7.74 \; m_e$. Since the $p\mu e$ system is below that value, only a finite number of bound states can exist. This is a first indication that due to the weakness of the attractive dipole potential, the existence of resonant states is not assured.

For a more detailed investigation, we consider the full-three body dynamics and use a numerical diagonalization of the complex rotated~\cite{reinhardt1982} Schr\"odinger Hamiltonian in a variational basis set to obtain the energies and widths of resonances. Having in mind that our goal is to determine whether resonant states do exist, we start with the $p\mu\mu$ negative ion, where resonances below the $n=2$ threshold are known to exist, and then follow the position of the lowest resonant state (with ro-vibrational quantum numbers $v=L=0$) while decreasing the mass $m$ of the third particle -which we denote $x$ in the following- in small steps, from the muon down to the electron mass. We used the perimetric coordinates (denoted $x,y,z$) and Sturmian wave functions
\begin{equation}
\begin{array}{@{}l}
\displaystyle \chi_{n_x,n_y,n_z}^{\alpha,\beta} (x,y,z) = (-1)^{n_x + n_y + n_z} \sqrt{\alpha\beta^2} \\
 \displaystyle\hspace{5mm} \times L_{n_x}\left(\alpha x\right) L_{n_y} \left(\beta y\right) L_{n_z} \left(\beta z\right) e^{-(\alpha x+\beta y+\beta z)/2}
\end{array}
\end{equation}
where $n_x,n_y,n_z$ are non-negative integers and $L_{n}$ the Laguerre polynomials. The main idea of this method~\cite{pekeris1958} is to take advantage of the dynamical symmetries of the three body Coulomb problem; in this basis set the Hamiltonian has strict coupling rules, leading to a sparse-band matrix which can be diagonalized in fewer operations with respect to a "full" matrix, maintaining better numerical stability~\cite{kilic2004,hilico2000}. This choice is well suited for the present study, in particular since the small number of variational parameters (the two length scales $\alpha^{-1}$ and $\beta^{-1}$) makes it easy to readjust them for each value of $m$.

\begin{figure}[h]
\begin{center}$
\begin{array}{c}
\includegraphics[width=8cm]{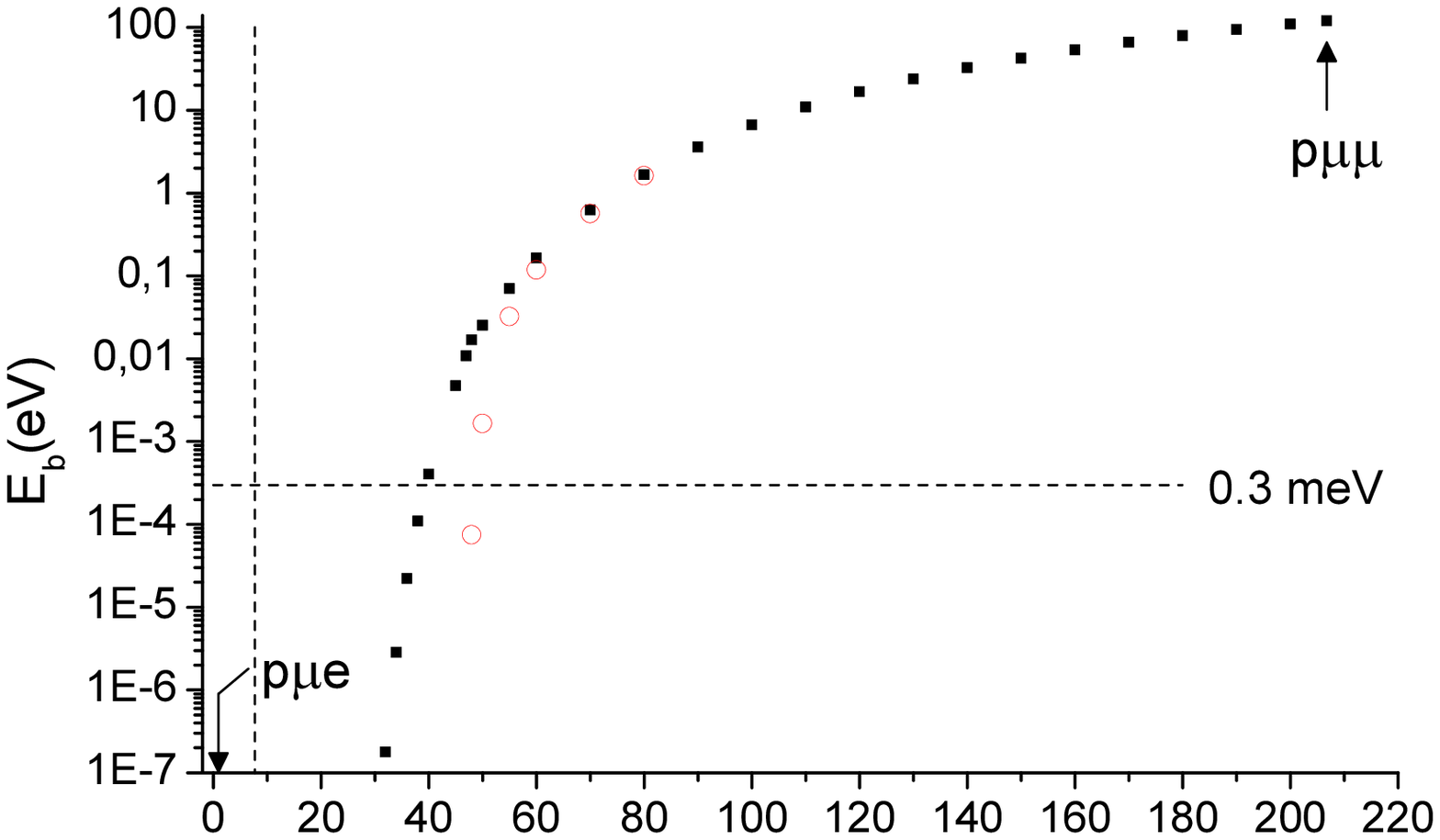} \\
\includegraphics[width=8cm]{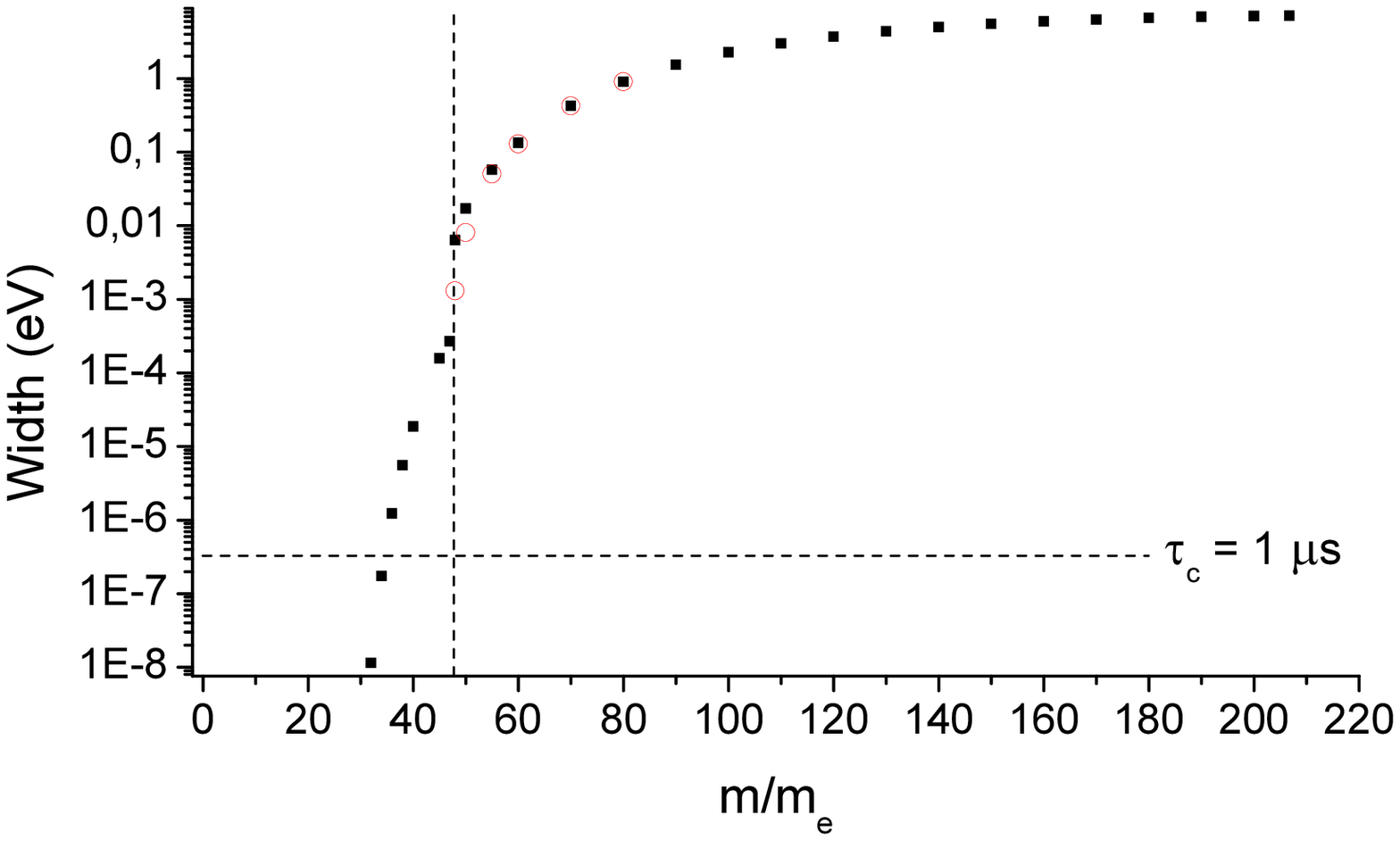}
\end{array}$
\end{center}
\caption{Binding energy (top) $E_b = E(p\mu)_{n=2} - E$ and Coulomb width (bottom) of the lowest $^1$S$^e$ resonance of the $p^+ \mu^- x^-$ system, as a function of the mass $m$ of the third particle $x$. Squares: pure Coulomb Hamiltonian, circles: Coulomb Hamiltonian with an added Yukawa potential adjusted to reproduce the 2S-2P Lamb shift. For the latter, the binding energy is with respect to the {\em shifted} 2S threshold. On the top graph, the horizontal line recalls the order of magnitude of the discrepancy between theory and experiment in the proton radius experiment, while the vertical line at $m/m_e = 7.74$ signals the minimal value of $m$ to retain an infinite series of resonant states in the adiabatic approximation. On the bottom graph, the horizontal line shows the value of the width corresponding to a Coulomb lifetime $\tau_c = 1 \; \mu$s, and the vertical one at $m/m_e = 47.67$ shows the crossing point between the $p\mu(n=2)$ and $px(n=1)$ thresholds.}
\label{pmue}
\end{figure}

Our results are summarized in Fig.~\ref{pmue}. Numerical accuracy is better than 10$^{-9}$~eV, so that even the smallest width, lying in the 10$^{-8}$~eV range, is obtained with two significant digits. The point at the extreme right corresponds to the $p\mu\mu$ negative ion; its binding energy $E_b = 120.4716427$~eV and width $\Gamma = 7.146063657$~eV (obtained with the 2010 CODATA values of the proton and muon masses~\cite{codata}), improve on the values reported in~\cite{ho1979}, $E_b = 120.39$~eV and $\Gamma = 7.12$~eV. As expected, when the mass of the third particle decreases, its orbit around the $p\mu(n=2)$ core becomes larger and its binding energy decreases.

Coulomb widths are shown mainly for illustration since the focus is on the {\em existence} of quasibound states. They decrease together with $E_b$ with decreasing $m$, because the system comes closer and closer to the adiabatic limit where the resonant state tends to a true bound state. The sharp drop between $m = 48 \, m_e$ and $m = 47 \, m_e$ is due to a crossing of the $p\mu(n=2)$ and $px(n=1)$ thresholds which occurs at $m = m_p m_\mu / (4m_p + 3 m_\mu) \simeq 47.67 \, m_e$. Below that value, the $px(n=1)$ threshold is above $p\mu(n=2)$, so that one of the channels for Coulomb deexcitation becomes forbidden, and the width decreases.

No calculations were performed below $m = 32 \, m_e$, where larger computer resources would be needed to converge the resonance width with sufficient accuracy. However, the results of Fig.~\ref{pmue} are amply sufficient to demonstrate that for $m = m_e$ the binding energy of a resonant state would be too small by many orders of magnitude to resolve the $0.31$~meV discrepancy in the muonic hydrogen experiment.

\subsection{Inclusion of QED effects} \label{subsec-yukawa}

An important ingredient is still missing in our treatment, that is the Lamb shift which lifts the degeneracy of the $n=2$ level, leading to a zero static dipole. The dipole potential in $-A/R^2$ is thus replaced by an induced dipole potential decaying like $-\alpha_0 /2R^4$ where $\alpha_0$ is the polarizability of the 2S (or 2P) state. As a consequence, when resonances come closer and closer to the threshold, they experience a more and more weakened attractive potential. The infinite dipole series obtained for $1/R^2$ potentials becomes truncated since a $1/R^4$ potential only admits a finite number of bound states~\cite{wallenius1996,lindroth1998}. It is therefore clear that the QED shift of energy levels does not put into question the above conclusion, since it weakens the attractive potential and can only favor the disappearance of weakly bound resonant states.

In order to show this more quantitatively, we represent QED corrections by an effective potential added to the three-body Coulomb Hamiltonian before diagonalization (the perturbative approach is not valid here due to the mixing of $2s$ and $2p$ orbitals). The most natural choice would be the Uehling vacuum polarization potential which accounts for more than 99\% of the $\mu p$ Lamb shift. This approach was used to obtain the energy shifts of $dt\mu$ resonances using the stabilization technique with a real scaling parameter~\cite{wallenius1996}. However, that method would not yield precise enough results here because the states under consideration have a large width that is not negligible with respect to their binding energy. In principle, the complex rotation method allows for higher precision, but numerical experimentation suggests that the Uehling potential is not dilation analytic (see e.g.~\cite{reed} for a definition of this term) and thus cannot be easily combined with the complex rotation method. For this reason, we ''mimic'' the proton-muon Uehling potential by a Yukawa potential $V_{eff}(r) = - \lambda e^{-\alpha r}/r$. First, we determined the exact shifts of the 2S and 2P levels in $\mu p$ induced by the Uehling potential by diagonalization of the two-body Coulomb+Uehling Hamiltonian: $\Delta E(2P) = -14.5789$~meV and $\Delta E(2S) = -219.7374$~meV~\cite{2p-level}. The deduced contribution to the Lamb shift, $\Delta E(2P)- \Delta E(2S) = 205.1584$~meV is in exact agreement with the result of~\cite{jentschura2011a}. Then, the parameters $\alpha = 492.86$ and $\lambda = 2.050982 \; 10^{-3}$ (in atomic units) were adjusted so that the diagonalization of the two-body Coulomb+Yukawa Hamiltonian yield the same exact shifts.

Going back to the three-body problem, the perimetric coordinates method loses its point here, because the added potential, having no selection rules, is represented by a ''full'' matrix. We thus used another variational expansion with Hylleraas coordinates $r_1,r_2,r_{12}$ and exponential basis functions
\begin{equation}
\chi_n^{1(2)} (r_1,r_2,r_{12} ) = \mbox{Re(Im)} \bigl[e^{-\alpha_n r_{12}-\beta_n r_1-\gamma_n r_2}\bigr]
\end{equation}
with pseudorandom complex exponents $\alpha_n, \beta_n, \gamma_n$~\cite{korobov2000}. In order to test the validity of the Yukawa approximation, we computed the energy of the lowest resonances of $dt\mu$ below the $n=2$ threshold. Agreement with calculations using the exact Uehling potential~\cite{wallenius1996} is obtained at the meV level (see Table~\ref{yukawa}). The approximation is expected to work even better for more weakly bound resonances, since the outer particle's orbit lies farther from the $\mu p(n=2)$ core and will be less sensitive to the short-range details of the proton-muon interaction potential. The main approximation in our approach is the neglect of fine and hyperfine structure (e.g. the 2S$_{1/2}$ hyperfine splitting is 22.8 meV). This is not expected to strongly modify our results, since the polarizabilities of individual hyperfine levels differ from the structureless case by less than 2 percent. We estimate that QED-induced changes in binding energies and widths are obtained with a relative accuracy of a few percent.
\begin{table}
\begin{tabular}{@{\hspace{1mm}}c@{\hspace{1mm}}@{\hspace{1mm}}c@{\hspace{1mm}}@{\hspace{1mm}}c@{\hspace{1mm}}@{\hspace{1mm}}c@{\hspace{1mm}}@{\hspace{1mm}}c@{\hspace{1mm}}}
\hline\hline
\multicolumn{5}{c}{$dt\mu$} \\
\hline
$v$ & $E_b$ (eV)~\cite{kilic2004} & $E_b^Y (eV)$  & $E_b^U$ (eV)~\cite{wallenius1996} \\
\hline
0 & 217.889 8 & 217.828 9 & 217.829 \\
\hline\hline
\multicolumn{5}{c}{$p\mu x$} \\
\hline
$m/m_e$ & $E_b (eV)$ & $E_b^Y$ (eV) & $\Gamma$ (eV) & $\Gamma^Y$ (eV) \\
\hline
80 & 1.673 0  & 1.611 0 & 0.907 7 & 0.907 1 \\
70 & 0.623 5  & 0.567 1 & 0.423 9 & 0.422 6 \\
60 & 0.165 2  & 0.118 2 & 0.133 5 & 0.129 1 \\
55 & 0.070 2  & 0.032 2 & 0.057 7 & 0.050 1 \\
50 & 0.025 3  & 0.001 6 & 0.017 1 & 0.008 1 \\
48 & 0.016 8  & 7(6).10$^{-5}$  & 0.006 4 & 0.001 3 \\
\hline\hline
\end{tabular}
\caption{Binding energies of resonant states of $dt\mu$ (top) and $p\mu x$ (bottom) below the $\mu p$(2S) threshold. $E_b$ is obtained with the pure Coulomb Hamiltonian, $E_b^Y$ with an added Yukawa potential (see text), and $E_b^U$  (for $dt\mu$ only) with the exact Uehling potential. Resonance positions are given with respect to the {\em shifted} 2S atomic threshold. For $p\mu x$, the corresponding widths $\Gamma,\Gamma^Y$ are also given.\label{yukawa}}
\end{table}

Results are shown in Table~\ref{yukawa} and plotted in Fig.~\ref{pmue} (open circles). The drop of binding energy ranges from about 62 meV for $m = 80 \, m_e$ down to a few meV for $m = 50 \, m_e$. This is consistent with the picture of a resonant state as a $\mu p$ atom in a 2S orbital interacting with the field of the distant $x$ particle. Indeed, the 2S-2P mixing decreases as $x$ becomes more loosely bound so that the shift of the resonance tends to that of the 2S level (220 meV). Already for $m = 48 \, m_e$, the binding energy is in the 0.1 meV range. These results confirm that the $p\mu e$ system can have no resonance that would be sufficiently shifted with respect to the 2S atomic level to explain even a small fraction of the 0.3 meV discrepancy.

\section{Molecular states} \label{sec-ppmu}

As mentioned in the Introduction, metastable states of the $pp\mu$ molecule are known to play a role in the muonic hydrogen deexcitation cascade~\cite{wallenius2001,pohl2006}. However, their possible involvement in the observed transition signal is highly constrained by experimental data~\cite{pohl2010}. Firstly, the observed 18~GHz-wide transition cannot be linked to a photo-association (or photo-dissociation) process, which would lead to a much broader spectrum. Secondly, two hyperfine components were observed, and the second one~\cite{antognini} was found at the predicted frequency with the proton radius value deduced from the first one i.e. the measured line has a hyperfine structure that closely matches that of the 2S$_{1/2} \rightarrow$ 2P$_{3/2}$ atomic transition. Thus the only possible scenario involving molecular states is that of a transition between resonant states, a first one lying close to the $\mu p$(2S$_{1/2}$) threshold and a second one close to the $\mu p$(2P$_{3/2}$) threshold. A prerequisite to a resolution of the proton radius puzzle would be to find such a pair of states, connected by a dipole-allowed transition, with such binding energies that the transition would be blue-shifted by a fraction of meV with respect to the 2S$_{1/2}$-2P$_{3/2}$ atomic transition. One strong point of this hypothesis is that the lower state would have long enough Coulomb {\em and} radiative lifetimes (since its wave function mainly involves the 2S atomic orbital with extremely small admixture with 2P) to survive the $\sim 1\mu$s delay before the spectroscopy laser pulse; similarly, the upper state wave function being mainly of 2P character, the transition width could be close to the observed one.

\begin{figure}[h]
\begin{center}$
\begin{array}{cc}
\includegraphics[width=4.4cm]{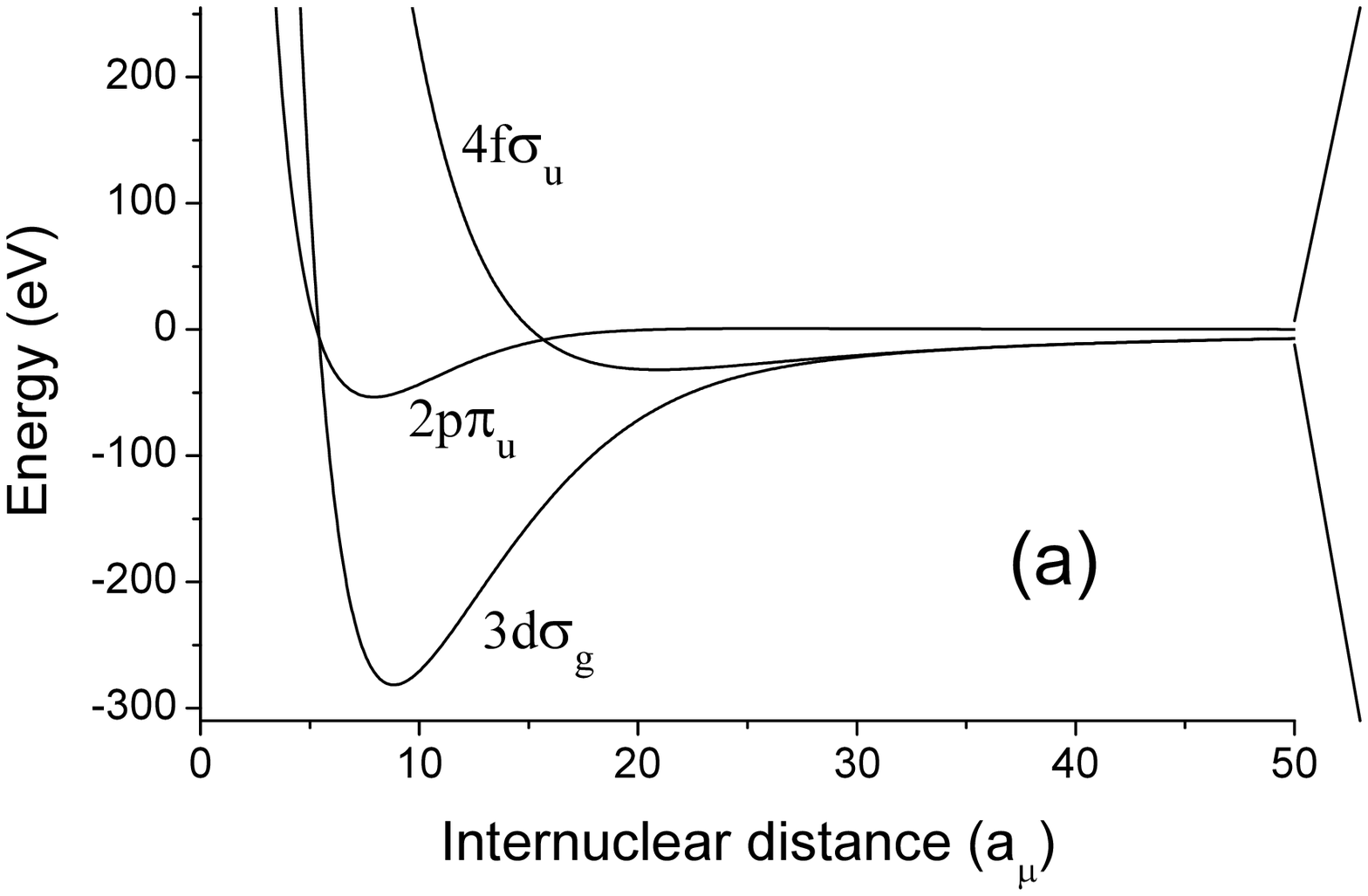} \includegraphics[width=4.1cm]{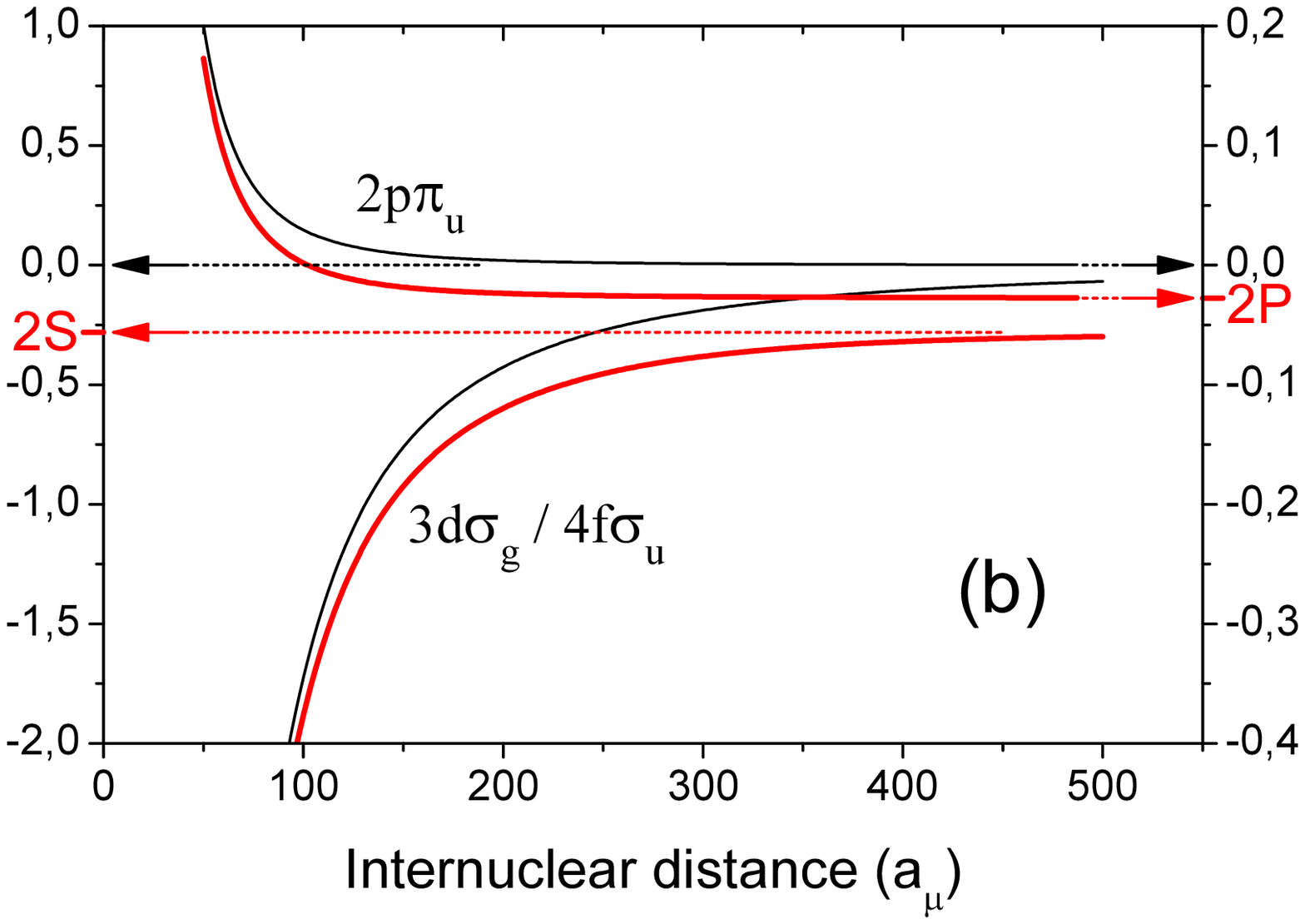}
\end{array}$
\end{center}
\caption{(a) Binding BO electronic energy curves below the $n=2$ threshold. Distances are in units of the muonic Bohr radius $a_{\mu} \simeq a_0 / 207$, and energies are with respect to the threshold. (b) Enlarged view of the region of large internuclear distances. The thin black lines are the continuation of the curves in (a), and converge to the unshifted $n=2$ threshold. The vertical scale on the left (right) correspond to the $\sigma$ ($\pi$) curves respectively. The $\sigma$ curves are indistinguishable at such large distances; their convergence is slower because they behave like $1/R^2$ instead of $1/R^4$. The thick red lines are obtained by adding a Yukawa potential $Y$ representing the proton-muon Uehling potential. The $2p\pi_u$ curve converges to the shifted 2P threshold shown on the right vertical axis. The $3d\sigma_g$ and  $4f\sigma_u$ curves converge to the shifted 2S threshold shown on the left axis. Since BO curves are drawn for infinite nuclear masses, for consistency the vacuum polarization shifts and associated parameters of $Y$ were calculated for a $\mu p$ atom with infinite proton mass: $\Delta E(2S) = -281.5896$ meV and $\Delta E(2P) = -27.5666$ meV yielding an exponent $\alpha = 443.7624$ and coefficient $\lambda = 1.919722 \; 10^{-3}$ for the Yukawa potential (see Sect.~\ref{subsec-yukawa}).}
\label{ppmu}
\end{figure}

There are three binding Born-Oppenheimer (BO) electronic curves below the $n=2$ threshold (see Fig~\ref{ppmu}a). The $3d\sigma_g$ and  $4f\sigma_u$ curves have a $1/R^2$ long-range behavior (dipole potential) and support infinite series of resonant states of symmetries $^1\!S^e$, $^3\!P^o$, $^1\!D^e$... and $^3\!S^e$, $^1\!P^o$, $^3\!D^e$... respectively~\cite{kilic2004,lindroth2003,shimamura1989}. The $2p\pi_u$ curve decays like $1/R^4$ (induced dipole potential) and supports a finite number of states with symmetries $^1\!P^o$, $^3\!P^e$, $^1\!D^o$, $^3\!D^e$...~\cite{hara1989,korobov1992,korobov1996}. The Lamb shift splits the $n=2$ threshold, and the key point is that both $\sigma$ curves converge to the $\mu p$(2S) threshold, while only the $2p\pi_u$ curve converges to the 2P threshold. This is clearly apparent in Fig~\ref{ppmu}b, which shows how the BO energies are modified by the Lamb shift at large internuclear distances. We obtained these curves by calculating the energy of an electron in the field of the nuclei for a grid of values of $R$, with a Yukawa potential added to the Coulomb potential in order to represent the Uehling potential, as described in Sect.~\ref{subsec-yukawa}. The electronic wave function was expanded on a variational set of exponential basis functions with pseudorandom exponents~\cite{tsogbayar2006}.
 
These considerations show that the upper level of the proposed molecular transition could only be one of the levels supported by the $2p\pi_u$ curve. These states have already been investigated in detail~\cite{hara1989,korobov1992,korobov1996}: one $^1\!P^o$ resonant state (with a binding energy $E_b = 13.57$ eV) and one ''anomalous parity'' $^3\!P^e$ bound state ($E_b = 13.54$ eV) have been evidenced, and it was shown convincingly in~\cite{korobov1992} that no other state can exist~\cite{shape-res}. We can thus conclude that the hypothesis of a molecular transition fails due to the lack of a suitable upper state.

\textbf{Conclusion.} We have refuted all reasonable hypotheses aiming to resolve the ''proton radius puzzle'' with the help of three-body physics. On the one hand, it was shown that the $p\mu e$ atom has no resonant states below the $\mu p(n=2)$ threshold; on the other hand, that the spectrum of the $pp\mu$ molecular ion leaves no room for a transition lying close to the 2S-2P atomic line, since there are no long-lived resonant states in the vicinity of the 2P threshold. However, the search for possible experimental artifacts in the muonic hydrogen experiment might not be over yet; as noted in~\cite{jentschura2011a}, formation and deexcitation of muonic hydrogen are very complicated processes~\cite{jensen2002} in which other many-body bound or quasibound states could play a role. Another result of this work is the demonstration of a convenient method to study the effect of QED level shifts on the position and width of few-body resonances in the framework of complex-rotated variational methods, using an adjusted Yukawa potential.

\textbf{Acknowledgments.} We thank V.I. Korobov for sharing his program for variational calculations of three-body systems with exponential basis functions, and V.I. Korobov, P. Indelicato, F. Nez, D. Delande, R. Pohl and F. Kottmann for helpful discussions.


\begin{thebibliography}{99}
\bibitem{2p-level} The 2P level shift is not negligible since the radius of the muon's orbit ($2 (m_e / m_r) a_0 \simeq a_0/92.7$) is of the same order as the characteritic length of the Uehling potential (the Compton length $\alpha \, a_0 \simeq a_0/137$).
\end{thebibliography}
\end{document}